# Operation of an optically coherent frequency comb outside the metrology lab

L. C. Sinclair,[1*] I. Coddington,[1] W. C. Swann,[1] G. B. Rieker,[1,2] A. Hati,[1] K. Iwakuni,[3] and N. R. Newbury[1*]

[1]*National Institute of Standards and Technology, 325 Broadway, Boulder, Colorado 80305*
[2]*Currently with the Department of Mechanical Engineering, University of Colorado, Boulder, Colorado 80309*
[3] *Department of Physics, Faculty of Science and Technology, Keio University, 3-14-1, Hiyoshi, Kohoku-ku, Yokohama 223-8522, Japan*
[*]*laura.sinclair@nist.gov, nnewbury@boulder.nist.gov*

We demonstrate a self-referenced fiber frequency comb that can operate outside the well-controlled optical laboratory. The frequency comb has residual optical linewidths of < 1 Hz, sub-radian residual optical phase noise, and residual pulse-to-pulse timing jitter of 2.4 - 5 fs, when locked to an optical reference. This fully phase-locked frequency comb has been successfully operated in a moving vehicle with 0.5 g peak accelerations and on a shaker table with a sustained 0.5 g rms integrated acceleration, while retaining its optical coherence and 5-fs-level timing jitter. This frequency comb should enable metrological measurements outside the laboratory with the precision and accuracy that are the hallmarks of comb-based systems.

## I. Introduction

Frequency combs can support cutting-edge measurements in areas that include optical clocks and oscillators, high-accuracy frequency and time transfer, precision molecular and atomic spectroscopy from the UV to THz regimes, high-accuracy LADAR, optical frequency metrology, arbitrary waveform generation, low phase noise microwave generation, and the search for exoplanets [1–4]. However, existing high-performance frequency combs are restricted to the well-controlled optical metrology laboratory while many of these applications require operation in less well-controlled indoor locations or even on mobile platforms. For example, a mobile dual-comb open-air-path spectrometer is an intriguing option for very precise and accurate trace gas sensing [5–15]. The advent of a new generation of mobile atomic clocks [16,17] and comb-based free-space optical frequency transfer [18] could permit precise timing networks, tests of relativity, and clock-based geodesy [19,20]. Finally, the precision and accuracy of comb-based LADARs are attractive for manufacturing and remote sensing in general [21–30]. Here we demonstrate a self-referenced fiber frequency comb that can operate in the field while retaining its optical coherence and 5 femtosecond-level timing-jitter. This comb should allow the full precision and accuracy of frequency combs to be harnessed for measurement problems outside of the optical metrology laboratory.

Soon after the first demonstration of Ti:Sapphire-based frequency combs [1,2], fiber-based frequency combs appeared as a strong alternative [31–36]. Their performance has quickly evolved to levels well beyond the requirements for most applications listed above. In addition, from their inception, the potential benefits of fiber-laser based frequency combs over solid-state combs were quite clear with regard to expense, robustness, and field operation [31–36]. However, *coherent* fiber-laser based frequency combs have not yet realized this potential for robust field operation; none have been operated successfully outside of the laboratory and under significant vibrations.

This limitation is connected with the basic design of most fiber frequency combs, which use a femtosecond fiber-laser mode-locked through nonlinear polarization rotation combined with single-mode fiber optics for the subsequent detection and stabilization of the offset frequency and optical comb teeth. In such a design, changes in fiber birefringence induced by strain, temperature, or humidity cause polarization wander that strongly affects both the mode-locking of the underlying femtosecond fiber-laser and the supercontinuum generation necessary for offset-frequency detection and required spectral coverage. For this reason, simple repackaging of the typical laboratory systems is inadequate to achieve an environmentally-robust, fieldable frequency comb. Several groups have discussed initial efforts on space-borne combs in two recent conference proceedings. KAIST has demonstrated an rf-stabilized femtosecond fiber-laser in satellite operation [37], and Menlo Systems has reported development of an rf-stabilized fiber frequency comb for operation on a sounding rocket [38]. The requirements for space-based operation include survivability at extreme accelerations, but operation in a relatively benign vibration environment. In addition, these systems rely on phase-stabilization to an rf clock both because optical clocks are not yet space-borne and for the utmost robustness. Here we explore different operational conditions required by the many potential applications of a frequency comb in a terrestrial environment, which require tight optical coherence and low timing jitter, achievable only by phase-stabilization to an optical reference. An optically coherent figure-8 type fiber comb has been operated with the femtosecond fiber-laser subsystem vibrated at g-levels [39], but its repetition rate was limited to 27-MHz, too low for a majority of comb applications, and the vibration insensitivity was limited to the fiber laser itself.

In this article, we report an optically-stabilized 200-MHz fiber-laser frequency comb operated in the strong vibrational environments typical of terrestrial platforms. The comb is phase-stabilized to an optical oscillator to achieve Hz-level residual linewidths and femtosecond-level residual timing jitter while in a moving vehicle and on a shaker table at up to 0.5 g rms. With modest vibration isolation, this comb should be fieldable on a variety of mobile platforms to allow comb-based measurements without sacrificing the high performance of coherent laboratory-based combs.

## II. Comb Design

The basic design of the fieldable coherent frequency comb is shown in Fig. 1a. It strongly leverages previous work, particularly from IMRA, on the use of polarization maintaining (PM) sub-systems, a semiconductor saturable absorber (SESAM) to achieve self-mode-locking and waveguide periodically-poled lithium niobate (PPLN) for efficient frequency doubling [33,40–45]. Below, we emphasize the design features critical for continuous coherent operation outside the laboratory, with additional details in the Appendix, before describing the comb performance.

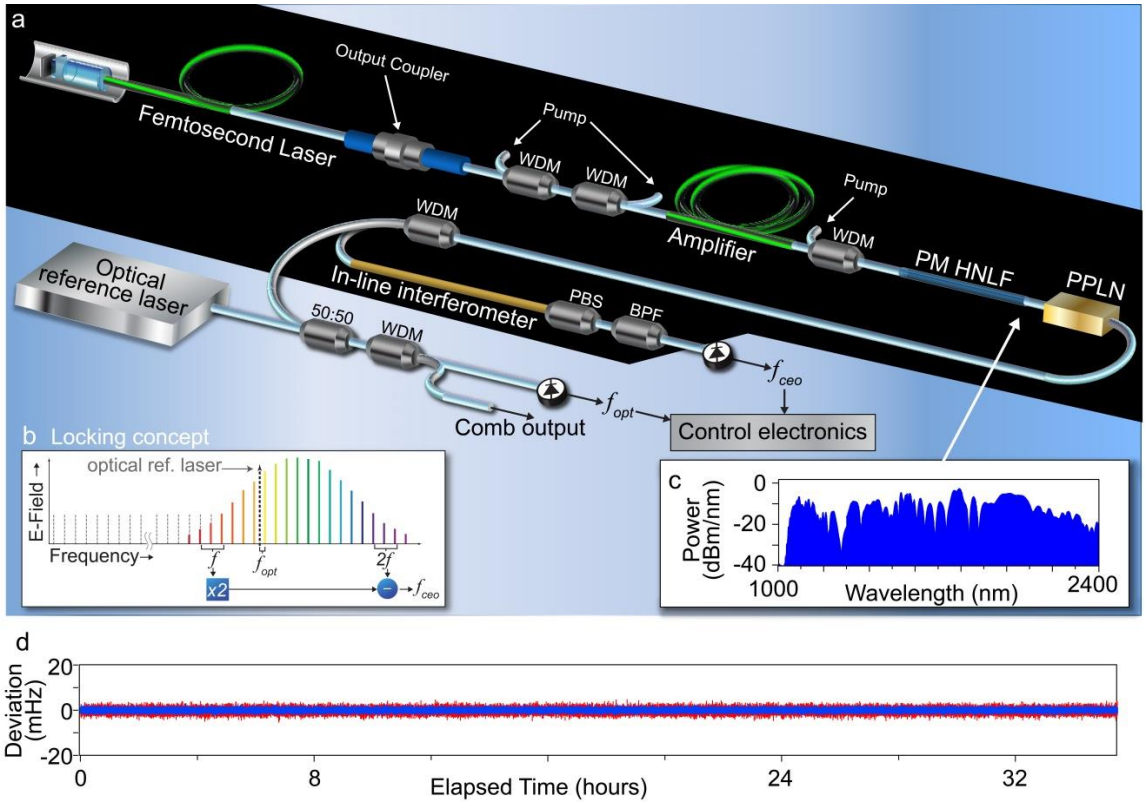

FIG. 1. Fiber Frequency Comb Design and Performance. (a) The fieldable coherent frequency comb is based completely on polarization maintaining (PM) fiber and telecom-grade fiber-optic components. The femtosecond laser is followed by an Er-fiber amplifier and then by a PM highly nonlinear fiber (HNLF) for spectral broadening. The carrier-envelope-offset frequency, $f_{ceo}$, is detected via the $f$-to-$2f$ approach [1,2] utilizing a waveguide periodically poled lithium niobate (PPLN) and PM "in-line" interferometer. The "optics package" that was shaken sits on the black background. WDM: Wavelength-Division Multiplexer, BPF: band-pass filter, PBS: polarizing beam splitter. (b) Schematic of phase-locking scheme. $f_{ceo}$ is phase-locked by feedback to the pump power. Optical coherence is established by phase-locking the optical heterodyne signal, $f_{opt}$, between a comb tooth and the optical reference laser through feedback to the cavity length. In the laboratory, the reference laser was a cavity-stabilized 1565 nm cw fiber laser, but for the tests in the vehicle and on the shaker table, the reference laser was a free-running 1535 nm cw fiber laser. (c) The octave-spanning spectrum output of the PM HNLF. (d) Counted $f_{ceo}$ (blue, standard deviation =0.62 mHz) and $f_{opt}$ (red, standard deviation =0.97 mHz) at 1 second gate time for over 35 hours without a phase slip.

Most importantly, the system is completely based on PM fiber and PM telecom-grade fiber optics components. All fiber connections are spliced or connectorized; there are no free-space sections, reducing sensitivity to vibration and alignment drift. The fiber-optic components are epoxied to the underlying Aluminum housing and the femtosecond fiber-laser is potted for vibration damping, but otherwise the housing includes no additional vibration damping.

The linear-cavity femtosecond Er-fiber laser is shown in the upper left of Fig. 1a. Previously, SESAM-based combs have achieved high repetition rates [43] and, with appropriate dispersion compensation, low noise [42,44,45]. Here, the SESAM design is attractive because it is compatible with a PM laser cavity and allows for true self mode-locking. The SESAM, which has 3% saturable loss and 6% non-saturable loss, is housed in a custom, telecom grade, micro-optic package designed to optimize the fluence at ~ 2 times the SESAM saturation value (20 μm diameter spot size), and provide fast-axis blocking. Additionally it prevents direct contact between the Er:fiber and the SESAM which can limit laser lifetime [43]. The laser's output coupler is provided by a broadband, 90% reflective, dielectric coating on the face of an FC/PC fiber connector [43]; this dichroic coupler transmits the 980-nm pump light. The 50 cm laser cavity contains ~ 28 cm of anomalous dispersion, highly-doped PM Er:fiber and ~ 22 cm of standard PM fiber, resulting in a round-trip cavity dispersion of ~ -0.02 ps$^2$, which supports soliton mode-locking. This simple femtosecond laser design produces output pulses with 10 nm of bandwidth, centered at 1567 nm, and 25 pJ power (5 mW). (See also the Appendix for a detailed description of the comb system.)

Despite this relatively modest output, an octave-spanning spectrum covering 1.05 μm to > 2.3 μm is generated. (See Fig. 1c). To achieve this, the laser output is amplified in 2 meters of normal-dispersion, erbium-doped PM-fiber; the resulting 1.5 nJ pulses are then compressed in fiber to 70 fs before entering a 20-cm PM highly non-linear fiber (HNLF) [46]. Following the HNLF, a fiber-coupled waveguide PPLN [47] doubles the supercontinuum light at 2128 nm, which is heterodyned with the fundamental supercontinuum light at 1064 nm to detect the carrier-envelope-offset frequency, $f_{ceo}$. In previous fiber combs, temporal overlap of these 1064-nm signals was achieved by tuning the lengths of the HNLF and single-mode fiber before the PPLN [33]. However, the long (> 10 cm) fiber leads of the fiber-coupled waveguide PPLN prohibit this approach and instead we implement an "in-line" PM-fiber interferometer after the waveguide PPLN that exploits the differential delay between fast and slow PM-fiber axes to cancel the relative group delay of the two 1064-nm signals.

The resulting $f_{ceo}$ signal has 35 dB SNR in a 300 kHz bandwidth, possibly benefiting from low-noise supercontinuum generation in the PM HNLF [44]. However, it exhibits significant free-running phase noise due to Gordon-Haus jitter resulting from the relatively large net cavity dispersion [35,45,48]. This noise is reduced through phase-compensated feedback [49] to the pump power at a 150 kHz bandwidth. Finally, optical coherence requires tight phase locking to a cw reference laser via high-bandwidth feedback to the laser cavity length. A 90 kHz feedback bandwidth is provided by a small (2 mm cube) piezo-electric transducer (PZT) attached directly to the fiber-laser cavity with careful damping of any fiber resonances through clay or potting compounds.

### III. Laboratory Performance

Figure 1d shows that both $f_{ceo}$ and $f_{opt}$ are free of phase slips for over 35 hours, eventually limited by actuator dynamic range or occasional electrical transients. Figure 2 characterizes the comb's optical coherence. This coherence is evident in the single coherent peak in the optical heterodyne signal between the comb and the 1565 nm reference laser, as well as in the coherent peak for the "out-of-loop" heterodyne signal against a second cavity-stabilized laser at 1535 nm. The integrated phase noise is 2.9 rad for $f_{ceo}$ and is 0.22 rad for $f_{opt}$ from 5 MHz to 1 Hz, with values extrapolated to Nyquist of 2.9 rad and 0.23 rad, respectively. The equivalent pulse-to-pulse timing jitter is calculated based on the phase noise at these two lock points divided by the (angular) optical frequency separation, or $\sqrt{2.9^2 + 0.23^2}/(2\pi \times 192 \text{ THz}) = 2.4$ fs assuming uncorrelated noise.

From Fig. 2, the output is optically coherent around the optical lock point (1565-nm). If optical coherence is defined as an integrated phase noise below 1 rad, then the ~3 rad phase noise on the carrier-envelope-offset frequency translates to a projected coherence across a comb spectrum of ~140 THz, centered at the optical lock point. This covers the entire 1 to 2 micron optical output of the comb. For the mobile tests, the doubling of the $f_{ceo}$ phase noise halves this spectral range.

This performance is comparable to lab-based frequency combs [31–36]. More importantly, the performance meets the requirements of metrological applications. The in-loop stability of the counted comb frequencies is below one mHz, more than sufficient for use with optical clocks [50]. The 3-fs timing jitter and <1-Hz residual comb tooth linewidths are below the 10-20 fs timing jitter and Hz-level linewidths tolerated in precision LADAR, optical time transfer, and comb spectroscopy [10,18,22,51]. Lower $f_{ceo}$ phase noise, and therefore timing jitter, as needed for low phase-noise microwave-generation [52] should be possible with dispersion-managed laser cavities or higher feedback bandwidth on the $f_{ceo}$ phase-lock.

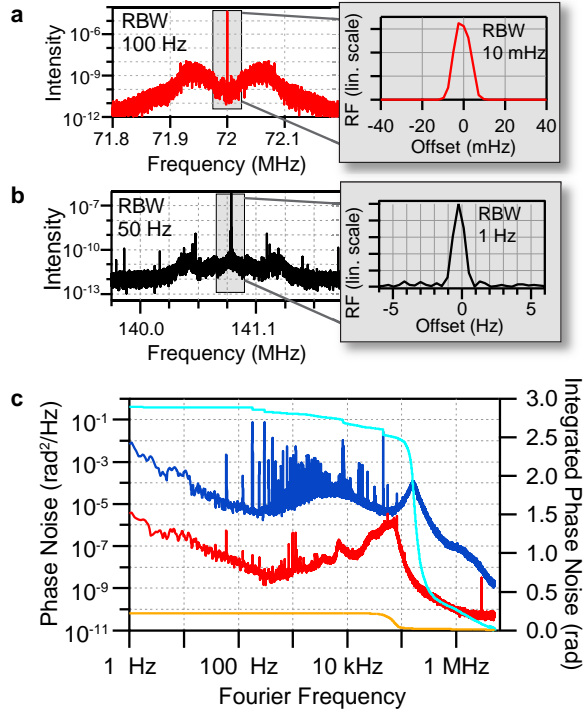

FIG. 2. Laboratory Performance (a) "In-loop" rf power spectrum of the comb tooth against the cw cavity-stabilized 1565 nm reference laser and (b) "out-of-loop" rf power spectrum against a second cavity-stabilized laser at 1535 nm, establishing the optical coherence of the comb. For the latter, the linewidth is instrument-limited at 1 Hz and the integrated phase noise is 0.7 rad from 5 MHz to 1 Hz. (The relative drift between the 1565 nm reference laser and this second cavity-stabilized laser limited the measurement to ~1 sec). RBW: resolution bandwidth. (c) "In-loop" phase noise power spectra and integrated phase noise for $f_{ceo}$ (blue and cyan) and $f_{opt}$ (red and orange). (The apparent phase noise increase below 100 Hz on both traces is due to the phase-noise analyzer, where the difference is due to the frequency division as described in the Appendix.)

## IV. Coherent Operation in a Moving Vehicle

To demonstrate this system's operation outside the laboratory, the frequency comb and test electronics were placed in a motor vehicle and powered by a small gasoline generator. As shown in Fig. 3, the phase noise of $f_{opt}$ was 0.3 rad, similar to in the laboratory, and phase noise of $f_{ceo}$ was 6.2 rad, twice the laboratory value due to the use of a different pump-diode current-controller which has a reduced feedback bandwidth. (This change was necessary due to the poor quality of the electrical power in the vehicle.) Residual pulse-to-pulse timing jitter was 5 fs. The phase-locks and the optical coherence were maintained while the vehicle was operated over significant dips, speed bumps, and a gravel road, with peak accelerations beyond 0.5 g. (Maintenance of the phase lock in the presence of both these optical frequency excursions and large temperature variations required occasional user adjustment of the overall pump current.) Note the absence of phase slips despite large optical frequency excursions of the free-running cw reference laser associated with the vehicle motion and temperature swings. In that regard, the coherence of the comb is certainly relative to the strongly varying cw reference laser frequency, but we emphasize this situation represents a worst-case limit to the comb performance. Even more stable comb operation would be achieved using a robust, mobile, cavity-stabilized cw reference laser such as in Ref. [16,17]. The supplemental movie associated with Fig. 3b shows excerpts from a 15-minute long drive with no phase slips, i.e. with fully coherent operation during the entire trip [53].

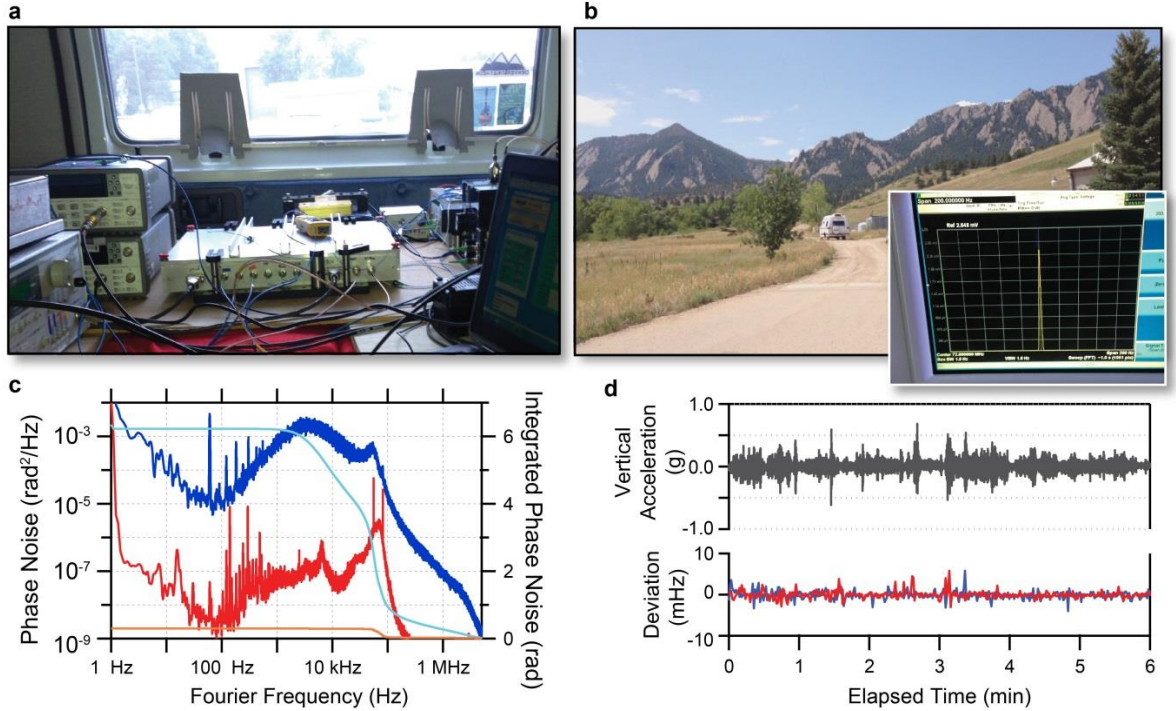

FIG 3: Coherent Comb Operation in Moving Vehicle. (a) Photograph of the comb inside of the vehicle. The Al box in the center contains the "optics package" (shaded black region in Fig. 1a). (b) Photograph of vehicle during comb operation. Inset: rf power spectrum of $f_{opt}$ while the vehicle travels over a speed bump. (RBW = 1.8 Hz, span = 200 Hz). (See also the Supplemental Movie [53].) (c) Phase-noise power spectral density and integrated phase-noise for $f_{ceo}$ (blue and cyan) and $f_{opt}$ (red and orange) with the vehicle and generator operating. (d) Deviation of the counted $f_{ceo}$ (blue, standard deviation = 1.3 mHz) and $f_{opt}$ (red, standard deviation = 0.96 mHz ) at 1 second gate time along with the vertical acceleration (grey) measured on the optics package while vehicle was in motion.

## V. Coherent Operation under Significant Vibrations

Vibrations in the vehicle were suppressed by its suspension, as they would be in any vehicle. To quantify the performance under stronger vibrations, the optics package (black region in Fig 1a) was mounted to a shaker table, as shown in Fig. 4. The vibration profile varied across the optics package due to resonances but covered from ~10 Hz to 1 kHz with a ~$1/f$ scaling and progressively increasing integrated rms values of 0.067g, 0.13g, 0.25g, 0.5g, and 1.3 g. (For the vibration spectrum at 0.035g, the shaker is off and background vibrations from various motors are seen primarily at 60 Hz and 120 Hz.) At low vibrations, the phase noise of $f_{opt}$ was ~ 0.5 rad and phase noise of $f_{ceo}$ was ~ 6 rad, due to strong electromagnetic interference in this noisier environment. Remarkably, as the vibration was increased beyond 1g, the $f_{ceo}$ phase noise was unchanged. (See Fig. 4e.) Because the $f_{ceo}$ phase noise dominates the timing jitter, the overall ~5-fs timing jitter is also independent of vibration level. The integrated phase noise on $f_{opt}$ increased roughly linearly with vibration level at 3.8 rad/g with a corresponding contribution to the timing jitter of 3.1 fs/g. This phase lock was unreliable beyond 0.5 g rms because of limitations in the PZT fiber-stretchers.

As the packaging of the optics system was minimal, this vibration test was considered as a test of the underlying optical design. Improved mechanical package design and vibration isolation would certainly further reduce vibration sensitivity of the optical phase lock. However, even with this simple packaging, the performance can be compared to the vibration profile of several platforms (beyond the moving vehicle for which successful operation was demonstrated in Section IV). Figure 4b compares the applied vibration profile at an integrated acceleration of 0.5 g (the highest vibration for which the $f_{opt}$ phase-lock was maintained) to the expected vibration levels for three different platforms assuming the frequency comb would be placed upon a typical air-mount (pneumatic isolator) with a natural frequency at 2.5 Hz and a 6% damping ratio [54]. Based on this comparison, the frequency comb so mounted should maintain successful operation on a ship, in a truck traveling across U.S. highways, and in a C-5 aircraft [55].

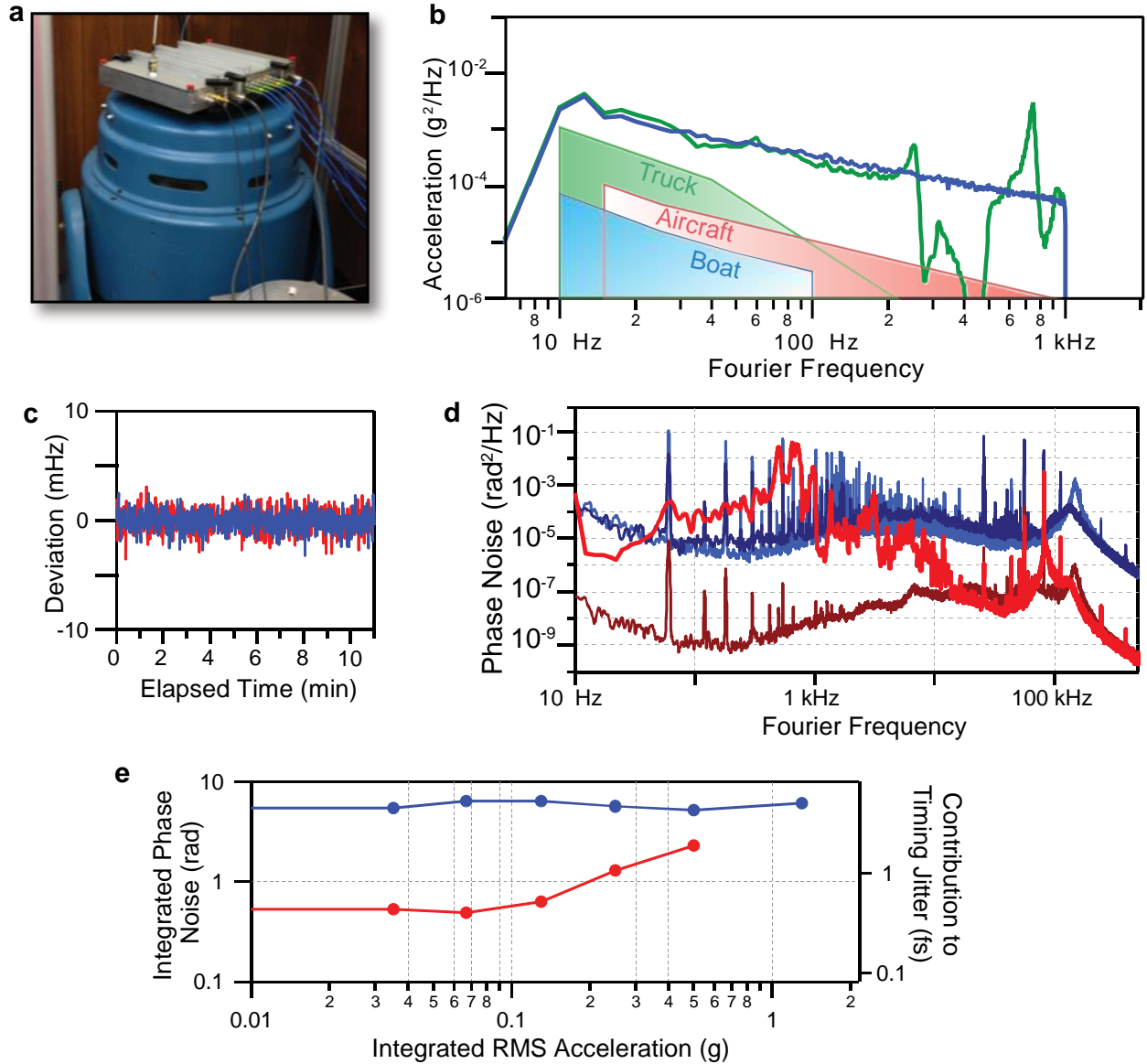

FIG 4: Coherent Operation Under Vibration (a) Optics package (Al box) mounted on the blue shaker table. (b) Applied $1/f$ vibration spectrum at the center of the Al box (blue line) and above the femtosecond laser location (green line) at an integrated value of 0.5g rms. In addition, the expected range of vibrations from several different mobile platforms is shown assuming a typical air-mount under the optics package [54,55]. (c) Deviation of counted $f_{ceo}$ (blue, standard deviation = 0.84 mHz) and $f_{opt}$ (red, standard deviation =0.90 mHz) frequencies at 1 second gate time during vibration with the 0.50-g integrated profile. (d) Phase noise power spectral density for $f_{ceo}$ at 0-g rms (light blue) and 1.5-g rms (dark blue), and for $f_{opt}$ at 0-g rms (dark red) and 0.5-g rms (bright red). (e) Integrated phase noise (left axis) and contribution to pulse-to-pulse timing jitter (right axis) for $f_{ceo}$ (blue) and $f_{opt}$ (red) versus integrated rms acceleration.

## VI. Conclusions

We have demonstrated a fiber frequency comb spanning 1.05 μm to above 2.3 μm with a 200 MHz repetition rate. This frequency comb has residual optical linewidths of < 1 Hz, sub-radian residual optical phase noise, and residual pulse-to-pulse timing jitter of 2.4 - 5 fs, when locked to an optical reference. To demonstrate the potential of this comb, the fully phase-locked frequency comb has been successfully operated in a moving vehicle with 0.5 g peak accelerations and on a shaker table with a sustained 0.5 g rms integrated acceleration, while retaining its optical coherence and 5-fs-level timing jitter. This frequency comb will enable metrological measurements outside the laboratory with the precision and accuracy that are the hallmarks of comb-based systems.

## Acknowledgments

This work was funded by the DARPA PULSE and QuASAR programs and by the National Institute of Standards and Technology (NIST). We acknowledge donation of the critical PM HNLF fiber by Masaaki Hirano of Sumitomo Electric. We acknowledge helpful discussions with David Leibrandt and Franklyn Quinlan and assistance from Craig Nelson, Lindsay Sonderhouse, Don Rieker, Marilyn Rieker and Marla Dowell.

## VII. Appendix: Detailed design of the fieldable coherent frequency comb

Figure 5 shows a more detailed design of the fieldable coherent frequency comb. The femtosecond fiber laser cavity consists of the SESAM micro-optic for mode-locking and fast-axis blocking, ~ 50 cm of fiber, and a dielectric coated fiber connector, which serves as the input/output coupler. The femtosecond laser is followed by an Er-fiber amplifier and then by a PM highly nonlinear fiber (HNLF) for spectral broadening. The optics package, shaded grey region in Figure 5, is housed in an Al box with 29 cm x 42 cm x 5 cm outer dimensions. No particular effort was made to minimize the size of the box and future efforts could significantly reduce the size of the box and further decrease its sensitivity to vibrations.

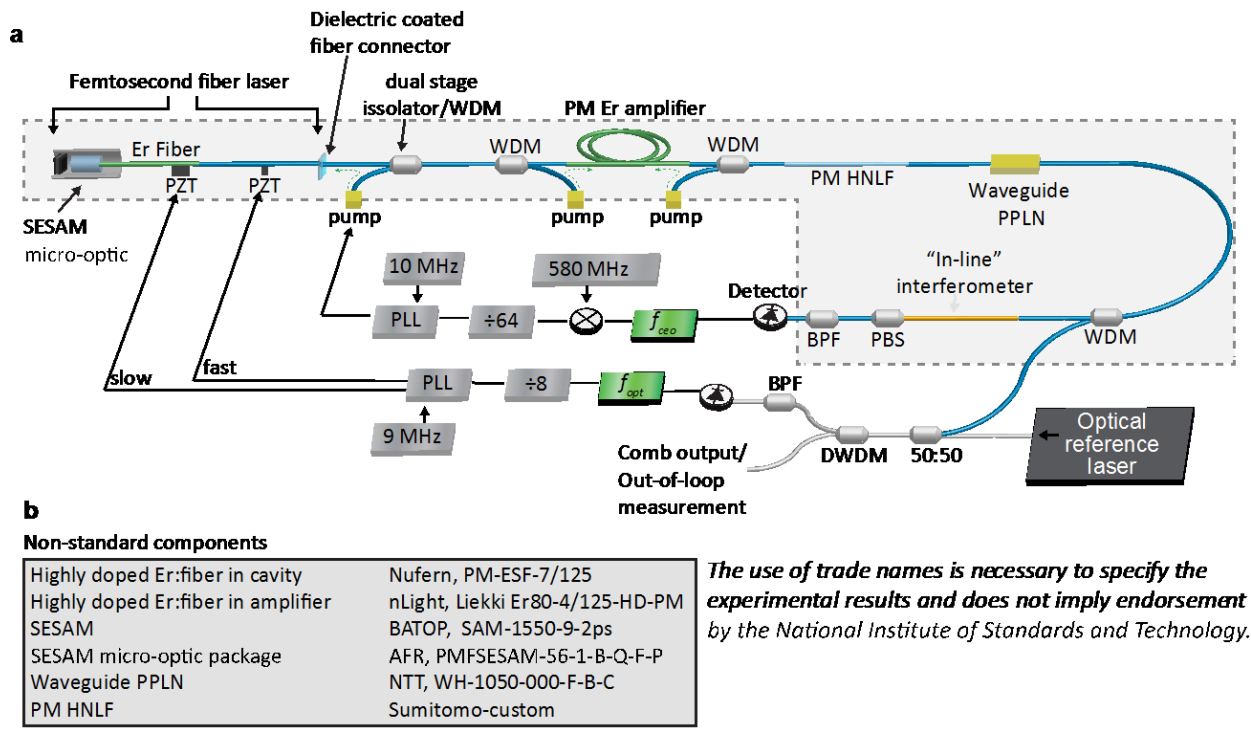

FIG. 5. (a) Fieldable coherent frequency comb design. A detailed discussion is provided in the text. The "optics package" that was shaken is shaded in light grey. PM fiber: solid blue lines, PM Er-doped fiber: solid green lines, WDM: Wavelength-Division Multiplexer, BPF: band-pass filter, PLL: phase-locked loop, PBS: polarizing beam splitter. (b) Table lists specific non-standard components used in the design.

$f_{ceo}$ is detected via the $f$-to-$2f$ approach [1,2], which requires the temporal overlap of the fundamental and frequency-doubled light. Immediately after the waveguide PPLN, the doubled and fundamental 1064 nm light are temporally separated by ~ 0.7 ps. We construct a PM-fiber "in-line" interferometer to overlap the pulses as follows. (See Fig. 6.) After the PPLN, a PM fiber is spliced to a second PM fiber section at a 45 degree angle to project the fundamental and doubled light onto both the fast and slow axes of this section. The light projected onto the slow axis experiences a delay relative to the light on the fast axis. After an appropriate PM fiber length (~1.6 m), the differential delay between the fast and slow axes compensates the 0.7 ps delay between the input fundamental and doubled light. We then project all the light back to the original basis with a second splice at a 45 degree angle. A polarizing beam splitter (PBS) selects the light on the slow axis and removes any interference from the light on the fast axis. Although the effective insertion loss is ~ 6 dB, the $f_{ceo}$ signal is strong enough for a robust lock (SNR > 35 dB in a 470 kHz resolution bandwidth), as shown in Fig. 6b. Moreover, the interferometer is "in-line", with almost all fiber length variations common-mode, so that there is negligible added excess phase noise to $f_{ceo}$.

To phase lock the comb, $f_{ceo}$ and $f_{opt}$ are detected on 100 MHz detectors, band pass filtered and amplified. To accommodate the higher fluctuations of $f_{ceo}$, it is mixed up in frequency to 640 MHz, divided by 64 and phase locked to a 10 MHz signal through feedback to the pump diode power. The $f_{ceo}$ response versus pump power is 52 MHz/mW and a phase lead [49,56] increased the feedback bandwidth to 150 kHz. Optical coherence is established by phase-locking the optical heterodyne signal, $f_{opt}$, of a comb mode and optical reference laser through feedback to the cavity length. $f_{opt}$ is divided by 8 to increase the locking range and phase locked to a stable 9 MHz rf signal through feedback to a small (high bandwidth) PZT and a larger (high dynamic range) PZT stack. A resister in series with the larger, PZT stack rolls off the response at 200 Hz to prevent resonances. In the laboratory, the reference laser was a cavity-stabilized 1565 nm cw fiber laser, but for the tests in the vehicle and on the shaker table, the reference laser was a free-running 1535 nm cw fiber laser. The comb output was used for the second heterodyne signal with an additional cavity-stabilized laser for the laboratory out-of-loop measurement reported in Figure 2.

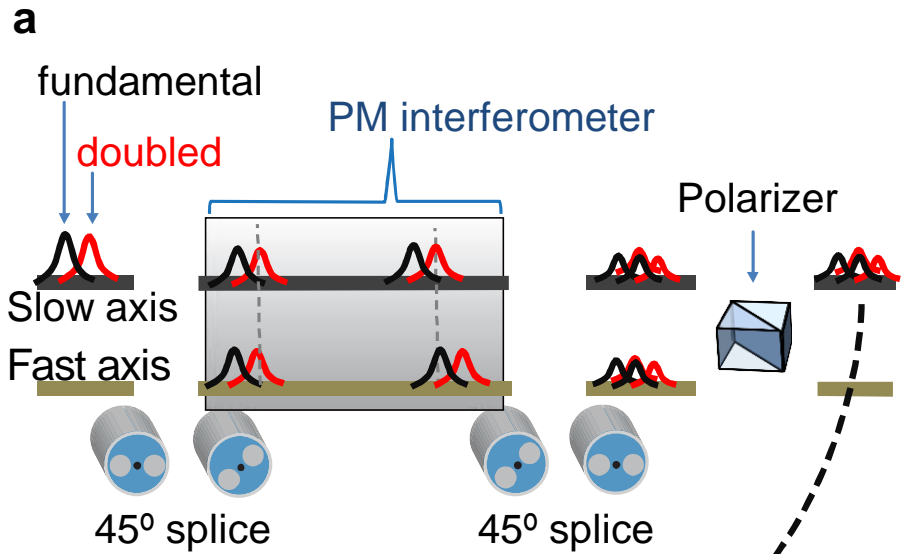

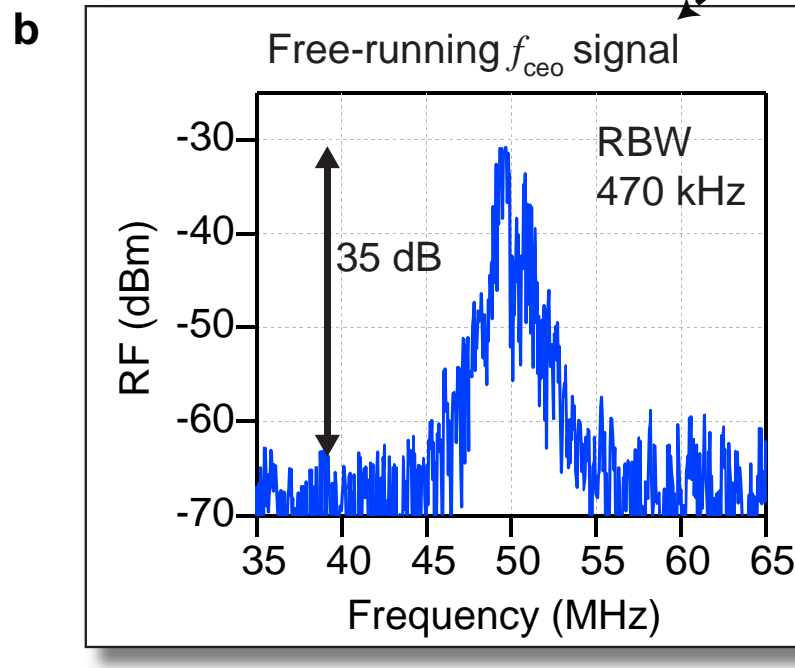

FIG. 6. Polarization-maintaining, in-line, fiber interferometer.